\documentstyle[12pt]{article}
\input epsf.tex
\input fps.sty
\def\fpsangle{0}
\newcommand{\pat}{\partial}
\newcommand{\be}{\begin{equation}}
\newcommand{\ee}{\end{equation}}
\newcommand{\bea}{\begin{eqnarray}}
\newcommand{\eea}{\end{eqnarray}}

\newcommand{\Acal}{{\cal A}}

\newcommand{\Rcal}{{\cal R}}
\newcommand{\Scal}{{\cal S}}
\newcommand{\Lcal}{{\cal L}}

\newcommand{\half}{\frac{1}{2}}

\begin{document}

\baselineskip 16pt

\begin{titlepage}
\begin{flushright}
CALT-68-2148 \\
hep-th/9712013 \\
December 1997
\end{flushright}

\vskip 1.2truecm

\begin{center}
{\Large {\bf Short Distance Contributions to}}
\end{center}
\vspace{-0.8cm}
\begin{center}
{\Large {\bf Graviton-Graviton Scattering:}}
\end{center}
\vspace{-0.8cm}
\begin{center}
{\Large {\bf Matrix Theory versus Supergravity}}
\end{center}

\vskip 0.8cm

\begin{center}
{\bf Esko Keski-Vakkuri}$^1$ and {\bf Per Kraus}$^2$
\vskip 0.3cm
{\it California Institute of Technology \\
     Pasadena CA 91125, USA  \\
     email: esko or perkraus@theory.caltech.edu}
\end{center}

\vskip 2.2cm

\begin{center}
{\small {\bf Abstract: }}
\end{center}       
\noindent           
{\small 
We study graviton scattering in the presence of higher dimensional operators
-- particularly, $\Rcal^4$ -- arising from loop effects.  We find 
that the results do not correspond to any known terms in the effective action
of Matrix Theory, thus lending support to the idea that the finite N Matrix
Theory has no simple relation to supergravity with large compactification
radii.
}
\rm
\vskip 3.4cm

\small
\begin{flushleft}
$^1$ Work supported in part by a DOE grant DE-FG03-92-ER40701.\\    
$^2$ Work supported in part by a DOE grant DE-FG03-92-ER40701 and by the
DuBridge Foundation.
\end{flushleft}
\normalsize 
\end{titlepage}

\newpage
\baselineskip 14pt

\section{Introduction}

Matrix Theory \cite{BFSS} has been successful in reproducing certain features
of classical, long distance supergravity in eleven dimensions \cite{CJS}.  
Graviton
scattering has been checked at the level of one and two graviton exchange
\cite{DKPS,BFSS,BB,BBPT},
including the leading order spin dependence \cite{Harvey,MSS,PK}.
Scattering processes involving extended branes have also been considered, and
such processes have found a unified description in terms of the Matrix 
gauge theory \cite{ChepTsey,Us}.  

   What has not been probed, however, is the short distance structure of
supergravity.  Viewing supergravity as a low energy effective field theory,
the short distance structure is encoded by the presence of higher dimensional
operators whose coefficients are suppressed by inverse powers of the Planck
mass.  The precise values of the coefficients are not calculable within the
framework of the low energy theory; such a determination requires a fundamental
description which is valid at short distances.  Matrix Theory purports to be
such a description, and so in principle allows one to compute the precise form
of the low energy effective action. 
 
Independent of Matrix Theory, there are certain higher dimensional operators
in supergravity whose coefficients are known.  One loop results in string
theory \cite{GS}, along with duality symmetries \cite{GV,GGV,RT}, strongly 
support the appearance of the
term \cite{GV}
\be
\Lcal_{\Rcal^4} = \frac{\pi^2}{9 \cdot 2^7 \cdot \kappa^{2/3}} 
\int \! d^{11}x \, \sqrt{g} \, t_8t_8 \Rcal^4
\ee
where the operator $t_8t_8 \Rcal^4$ is defined in (\ref{e11}) as a particular 
contraction of four Riemann tensors.  From the eleven dimensional point of 
view, the operator arises from a one loop diagram with short distance cutoff of
order the eleven dimensional Planck length.

The term above introduces new vertices which contribute to 
graviton-graviton scattering.  For instance, there is a new four point vertex
which can contribute through the diagram in Fig. 1.

\def\fpsangle{270}
\begin{center}
\leavevmode
\fpsxsize 1.5in
\fpsysize 1.5in
\fpsbox{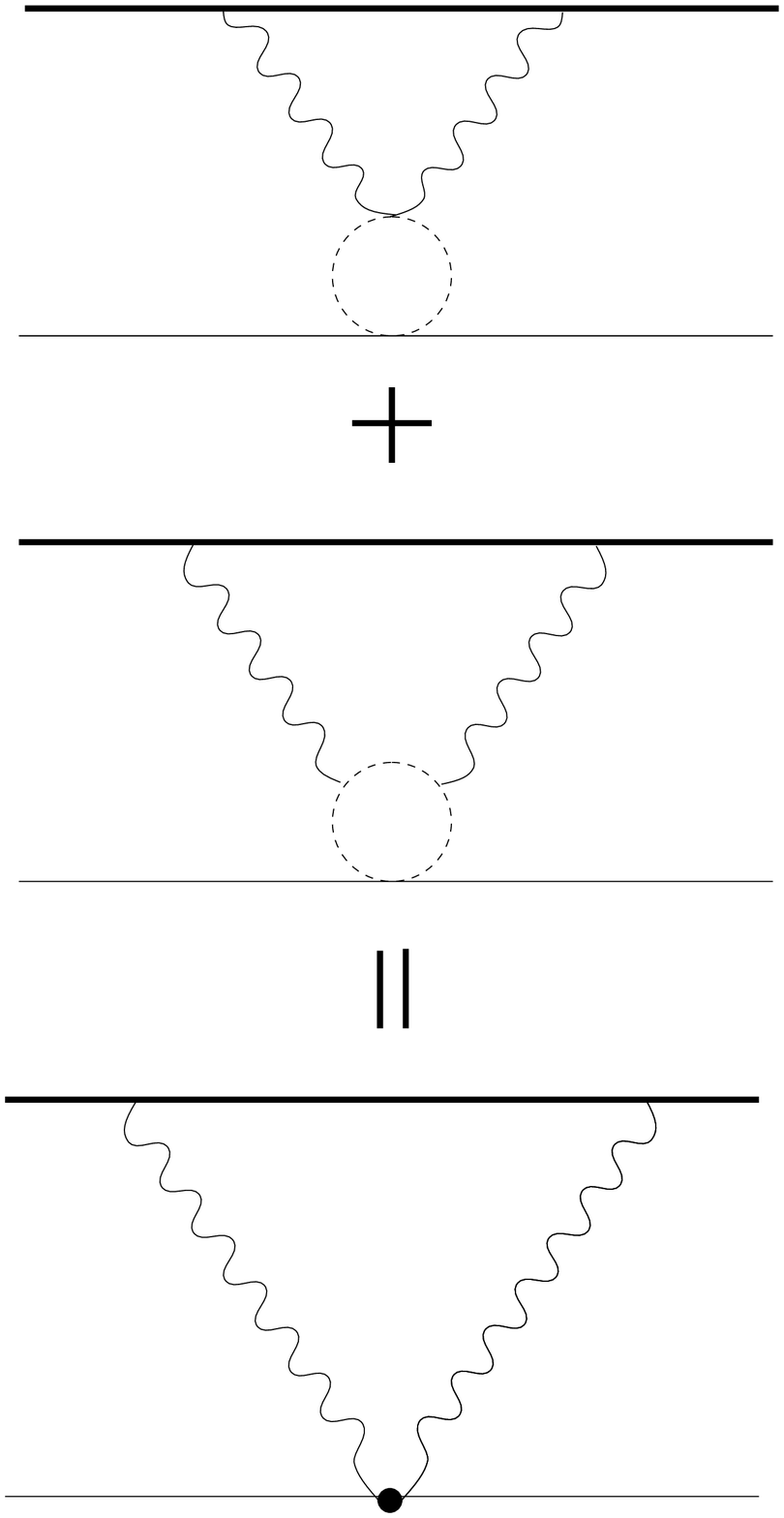}
\end{center}
\vskip 0.1 true cm
\begin{quotation}
\small
\noindent
{\bf Figure 1:}
Contribution to graviton-graviton scattering.  The thick and thin lines
represent the source and probe gravitons respectively (see text), and the
black dot in the leftmost diagram represents the  $t_8t_8 \Rcal^4$ operator. 
The two diagrams on the right hand side illustrate the origin of 
$t_8t_8 \Rcal^4$ as arising from the
divergent part of one loop diagrams.  The dotted lines represent 
 gravitons, anti-symmetric tensors, and gravitinos circulating in the loop.  
\end{quotation}
In sect. 3 we analyze this process by performing a generalized source-probe
calculation.  The result is expressed in terms of the effective action of the
probe, and takes the form
\be
L \sim \frac{N_{s}^2 N_p^3 v^8}{M^{24} R^7 r^{18}},
\label{sugres}
\ee
where $M$ is the eleven dimensional Planck mass, 
$N_{p}, N_{s}$ are the longitudinal momenta of the gravitons according
to $p_{-}=N/R$, and $v$ and $r$ are the relative transverse velocity and
separation.

On the Matrix Theory side, the effective action for the probe has the structure
\be
L_{l} = R M^{6-3 l} r^{4-3 l}f_{l}\left(\frac{v}{RM^{3}r^2}\right),
\label{lform}
\ee
where $L_{l}$ denotes the contribution from $l$ loop diagrams.  A 
particular two loop contribution is 
\be
L_{2}\sim \frac{v^8}{M^{24}R^{7}r^{18}}
\ee
which appears to match onto the term in (\ref{sugres}).  However, an $l$ 
loop diagram in Matrix Theory can give at most $l+1$ powers of $N$.  Thus
the supergravity contribution (\ref{sugres}), which has five powers of $N$,
cannot be reproduced from a two loop diagram in Matrix Theory.  In fact,
we see that there is no term in the expansion (\ref{lform}) which can reproduce
the supergravity result.

The appearance of the troublesome $N_p$ dependence in (\ref{sugres}) is easy
to understand by considering the structure of the  $\Rcal^4$ operator.  
$\Rcal^4$ has eight derivatives, and these act on the source and probe
variables.  The fact that (\ref{sugres}) goes as $r^{-18}$ means that 
four of the derivatives are acting on the source, since two powers of the
graviton propagator go as $r^{-14}$ if no derivatives act on them. The
remaining four derivatives act on the probe; each derivative yields a factor
of $N_p$, giving altogether $N_p^4$.  Finally, one has to divide by 
$p_- \sim N_p$ to obtain the proper normalization of states in the light cone
frame (or as is seen from the analysis in sect. 3).  This results in the
$N_p^3$ dependence of (\ref{sugres}). What is more
difficult to see is the numerical coefficient which multiplies the term in
(\ref{sugres}), and if it should happen to vanish then agreement with Matrix
Theory would be reached.  Therefore we are motivated to carefully compute 
this coefficient in sect. 3. We find that it is nonzero, indicating 
disagreement.   

Recently, there have appeared arguments \cite{goodnews} which  offer a 
derivation of the  equivalence between Matrix
Theory and supergravity with lightlike compactification.  However, as 
discussed in \cite{HP}, these arguments do not automatically apply to the
regime of supergravity which is being considered in this paper, namely large
compactification radii.  Thus discrepancies between finite $N$ Matrix Theory
\cite{Nfin} and the usual notion of supergravity cannot be 
ruled out, and indeed
such discrepancies have already appeared \cite{badnews}.

The remainder of this paper is organized as follows. In sect. 2 we review
the source-probe calculation of \cite{BBPT}, point out its limits of validity,
and show how it can be generalized.  We derive the effect of the 
$t_8t_8 \Rcal^4$ operator in sect. 3 and discuss 
the sorts of terms in the probe action
which can be generated. The calculation is somewhat involved due to the
fact that the effect we are looking at is polarization dependent, and so the
spin of the gravitons cannot be neglected even at leading order.  
 In sect. 4 we discuss the accuracy of our 
approach and elaborate on the roles of the anti-symmetric tensor
and gravitino, which have not appeared explicitly in our calculations.  
Finally, we comment on possible resolutions of the discrepancies.

\section{Graviton scattering in supergravity}

We begin by  reviewing  some results from previous work 
on graviton-graviton scattering. Ref. \cite{BBPT} studied the
scattering of two gravitons
in eleven dimensional supergravity with lightlike compactification. 
The null coordinate $x^-= x^{11}-t$ is taken to have periodicity
$2\pi R$. The lightlike momentum
$p_-$ is quantized in integer units of $1/R$, so $p_-=N_s/R,\ N_p/R$ for
the two gravitons. It is further assumed that $N_s$ is large enough 
that the first graviton can be considered to be a classical source, with the 
second graviton probing its gravitational field. That is, the scattering
has to be soft enough so that momentum transfer is negligible and there
are no recoil effects. The gravitational field of the source
is given by the metric
\be
 ds_{\rm bg}^2 = -2d\tau dx^- - h_{--} (dx^-)^2 - dx^2_{\perp}  \ ,
\label{bg}
\ee
where $\tau = x^+/2 = (x^{11}+t)/2$ and $x^i_{\perp}$ are the 9 transverse
coordinates. The graviton is located at $x_{\perp}=0$. A graviton with fixed
$x^-$ would produce the 
Aichelburg-Sexl metric \cite{AS}; however, here the graviton has a fixed
lightlike momentum so the metric must be averaged over the lightlike circle.
Upon averaging, the harmonic function $h_{--}$ takes the form 
\be
 h_{--} = \frac{15N_s}{2R^2M^9r^7} \ ,
\label{e3}
\ee
where $r^2 \equiv x^2_{\perp}$, 
$M$ is the Planck mass in eleven dimensions.  From now on
we will use the notation $h$ for the harmonic function $h_{--}$.

The authors of \cite{BBPT} find a point particle action for the probe graviton
    using the following strategy. They begin by writing the Lagrangian
for a massive scalar particle moving in the background (\ref{bg}). 
Then they point
out that the appropriate variational principle is one in which $p_-$ is held
fixed.  Thus instead of using the Lagrangian one should use the Routhian,
which is obtained by performing a Legendre transformation to eliminate 
$\dot{x}^-$ in favor of $p_-$.  Unlike the Lagrangian, the Routhian has a
non-trivial limit as $m\rightarrow 0$.  The (negative of the) Routhian is
\be
L'=L-p_{-}\dot{x}^-=\left[-m\sqrt{g_{\mu\nu}{\dot{x}}^{\mu}{\dot{x}}^{\nu}}
-p_{-}\dot{x}^- \right]_{m\rightarrow 0} = -p_{-}\dot{x}^-
\ee
where $\dot{x}^-$ is to be found from the equations of motion.  This procedure
produces the action
\begin{eqnarray}
  S = \int d\tau \ L'  &=&   
        \int d\tau \ \frac{p_-}{h} \{1-\sqrt{1-hv^2_{\perp}} \} \nonumber \\
    &=& \int d\tau \ \{ \frac{N_p}{2R} v^2_{\perp}
         + \frac{15}{16} \frac{N_pN_s}{R^3M^9} \frac{v^4_{\perp}}{r^7}
         + \frac{225}{64} \frac{N_pN^2_s}{R^5M^{18}} \frac{v^6_{\perp}}{r^{14}}
         + {\cal O}(\frac{v^8}{r^{21}}) \} \ ,
\label{Sprobe}
\end{eqnarray}
where $v^2_{\perp} = \dot{x}^i_{\perp}\dot{x}^i_{\perp}$. Note that
all terms are linear in the momentum of the probe $N_p$. 

This calculation, however, takes into account only the long distance
interactions 
arising from the Einstein term $\sqrt{g}\Rcal$ in the supergravity Lagrangian.
At shorter distances, there are corrections to the Lagrangian
involving higher order operators such as $\Rcal^4$. Such corrections will yield
additional terms\footnote{Some discussion can be 
found in \cite{BM}. An $r^{-27}$ contribution 
has been discussed in \cite{Ser}.} 
to (\ref{Sprobe}) and thus modify the graviton-graviton
amplitude.  
The corrections can be derived by computing Feynman diagrams, or
by a more convenient method which we now describe.  Essentially, one needs
to know how to pass from a field theory description of gravitons to a point
particle description, so that the field theory corrections -- like the presence
of the $\Rcal^4$ operator -- can be 
translated into corrections to point particle 
actions which are easily compared with Matrix Theory results.  The connection
between field and particle actions is provided by Hamilton-Jacobi theory, as
we now review in the simplified context of a scalar field theory.

Consider a massive scalar field
\be
 \Lcal = \half \int d^dx \sqrt{g} (\pat_{\mu} \phi \pat^{\mu} \phi 
     - m^2 \phi^2 ) \ .
\label{phiaction}
\ee
Neglecting derivatives of the metric, the field equation is 
\be
    g^{\mu\nu} \pat_{\mu}\pat_{\nu}\phi + m^2 \phi = 0 \ .
\ee
To pass to a point particle description, one uses the WKB form
\be
    \phi = e^{ip_{\mu}x^{\mu}}
\ee
which gives
\be
     g^{\mu\nu} p_{\mu}p_{\nu} - m^2 = 0\ .
\ee
Now the idea is to solve for $p_t(p^i,m)$ and to regard $H=-p_t$ as the
Hamiltonian of a point particle. Then one can work out the Lagrangian as usual
\be
 L = p_i \dot{x}^i + p_t 
   = -m\sqrt{g_{\mu\nu}\dot{x}^{\mu}\dot{x}^{\nu}} \ .
\label{ppaction}
\ee
One has thereby recovered the standard expression for the relativistic point
particle action.

The action (\ref{phiaction}) was of only second order in derivatives, but one can easily
add higher derivative terms, repeat the same procedure, and thus derive 
corrections to (\ref{ppaction}).  It is this method which we now follow
in the case of gravity.

\section{Calculations}

In this section we present the calculation of the effective action of a probe
graviton moving in the classical
background of a source graviton. We first show how to efficiently reproduce 
the result (\ref{Sprobe}) starting from the Einstein action.  Next, we extend 
the calculation to include the higher derivative $\Rcal^4$ operator.  
As we'll see,
the operator produces a term in the probe action that does not seem to 
arise from any straightforward Matrix Theory calculation.

We begin with the eleven dimensional Einstein Lagrangian
\be
    \Lcal = \frac{1}{2\kappa^2} \sqrt{g} \Rcal \ .
\label{e1}
\ee
We treat one of the two gravitons as a classical source which produces the
metric (\ref{bg}).
To this background, we add a small perturbation representing the presence
of the probe graviton
\be
 ds^2 = ds^2_{\rm bg} + \kappa f_{\mu \nu} dx^{\mu}dx^{\nu}
\label{e4}
\ee
Instead of considering simultaneously all of the components of $f_{\mu \nu}$, 
we will choose a 
single, fixed, transverse and traceless polarization\footnote{As discussed in 
Appendix A, polarization dependence shows up through terms in the probe
action which depend on derivatives of the harmonic function $h$.  The present
calculation correctly gives all the terms without derivatives.}. 
We take
\be
    f_{\mu \nu} = \left\{ 
    \begin{array}{l} f_{12} (\tau,x^-,x^3,\ldots ,x^9)\ , \ \mu \nu = 12,21 \\
                     0 \ \ \ \ \ \ \ \ \ \  , \  {\rm otherwise} 
                        \end{array} \right. \ .
\label{f12}
\ee
Then, substituting (\ref{e4}) into the Lagrangian yields the following terms of
second order in $f_{12}$
\be
    \Lcal = \frac{1}{4} \{ -2\pat_{\tau} f_{12} \pat_- f_{12} 
               - (\pat_{\perp} f_{12})^2
                  + h (\pat_{\tau} f_{12})^2 \} \ .
\label{e5}
\ee
Note that now $\perp$ denotes the indices $i=3,\ldots ,9$.
Now that we have a quadratic field action we can proceed as in sect. 2. 
The field equation is
\be
     (-2\pat_{\tau} \pat_- - \pat^2_{\perp} + h\pat^2_{\tau}) f_{12} = 0 \ .
\label{e6}
\ee
We use the WKB form for $f_{12}$, and substitute
\be
 f_{12} = e^{ip\cdot x}
\label{WKB}
\ee
into (\ref{e6}). This gives an equation for $-p_{\tau}$, 
the Hamiltonian of the probe, as a function of $p_-,p_{\perp},h$:
\be
  H = -p_{\tau} = \frac{p_-}{h} \{ \sqrt{1 + 
                 h p^2_{\perp}  /p^2_- } - 1 \} \ .
\label{e7}
\ee
The probe action is then given as the $\tau$ integral of the Routhian $L'$
constructed from (\ref{e7}):
\be
    L' = L - p_- \dot{x}^-
     = p_i \dot{x}^i + p_{\tau} 
             = \frac{p_-}{h} \{1- \sqrt{1 - hv^2_{\perp}} \} \ ,
\label{e8}
\ee
in agreement with the result in (\ref{Sprobe}), but now derived in a way that
readily admits generalization.  

Now, we repeat this analysis in the presence of an $\Rcal^4$ term arising from
 one-loop effects \cite{GS,GV}.
The Lagrangian is
\be
 \Lcal = \frac{1}{2\kappa^2} \sqrt{g} \Rcal + \frac{c}{\kappa^{2/3}} \sqrt{g}
 t_8t_8 \Rcal^4 
\label{e9}
\ee
where $c$ is a numerical coefficient argued to be \cite{GV}
\be
    c = \frac{\pi^2}{9\cdot 2^7} 
\label{e10}
\ee
and $t_8$ is an 8-index tensor, contracted with the Riemann tensor as follows:
\be
 t_8t_8\Rcal^4 = t^{\mu_1 \cdots \mu_8}_8 t_8^{\nu_1 \cdots \nu_8}
     \Rcal_{\mu_1 \mu_2 \nu_1 \nu_2} \cdots \Rcal_{\mu_7\mu_8\nu_7\nu_8} \ .
\label{e11}
\ee
For our perturbative considerations, we can follow the example of \cite{GV} 
and drop total derivative terms from (\ref{e11}); in this case the 8-index
tensor $t_8$ is given explicitly by 
\be
  t_8^{\mu_1 \cdots \mu_8} = \Scal \Acal \ \{ 24 \ \eta^{\mu_2\mu_3}
      \eta^{\mu_4\mu_5} \eta^{\mu_6\mu_7} \eta^{\mu_8 \mu_1}
     - 6\  \eta^{\mu_2\mu_3}
      \eta^{\mu_4\mu_1} \eta^{\mu_6\mu_7} \eta^{\mu_8 \mu_5} \} \ ,
\label{e12}
\ee
where $\Scal, \Acal$ are symmetrization and antisymmetrization operations
on the indices, such that  $t_8$ is antisymmetric in each 
pair of indices $\mu_1,\mu_2;\ \mu_3,\mu_4;\  \ldots$ and symmetric
under the interchange of any pair of these index pairs (see \cite{GSW}).  
These symmetry properties correspond to the symmetries of the 
Riemann tensor.

Next, we separate the total metric
into a flat contribution and a correction term:
\be
 ds^2 = 
      -2d\tau dx^- - dx^2_{\perp} -h(dx^-)^2 
 - \kappa f_{\mu \nu} dx^{\mu}dx^{\nu}
      = (\eta_{\mu \nu} + \Delta_{\mu \nu}) dx^{\mu}dx^{\nu} \ ,
\label{e13}
\ee
where
\be
  \Delta_{\mu\nu} = -h\delta_{\mu -}\delta_{\nu -} -\kappa f_{\mu \nu}. 
\label{e14}
\ee

 It would be an arduous task
to try to find all corrections to the Routhian of the probe arising from
the $\Rcal^4$ term.  We will instead focus on the leading term in the 
$1/r$ expansion, which goes as $r^{-18}$, and is sufficient to demonstrate
the apparent discrepancy with Matrix Theory.
Let us consider first what sort
of terms can give rise to $r^{-18}$ so that we know what we need to keep
track of and what to neglect. Recalling $h \sim r^{-7}$, we  quickly see
that that the combination 
\be
       \pat \pat h \pat \pat h \sim \frac{1}{r^{18}} 
\label{e15}
\ee
is the one that counts. Since $\Rcal_{\mu\nu\rho\sigma}$ and $\Rcal^4$ contain
at most second  derivatives of the metric, the terms of the 
type (\ref{e15}) are in fact the only ones that give $r^{-18}$ dependence.
Thus many complicated terms can be dropped.
For example, substituting $g_{\mu \nu} = \eta_{\mu \nu} + \Delta_{\mu \nu}$
into $\Rcal_{\mu\nu\rho\sigma}$ yields 
\be
    \Rcal_{\mu\nu\rho\sigma} = 
 -\half \{ \pat_{\mu} \pat_{\rho} \Delta_{\nu \sigma}
 - \pat_{\nu} \pat_{\rho} \Delta_{\mu \sigma} 
 - \pat_{\mu} \pat_{\sigma} \Delta_{\nu \rho}
 + \pat_{\nu} \pat_{\sigma} \Delta_{\mu \rho} \} + \cdots
\label{e16}
\ee
where $+ \cdots$ denotes irrelevant terms which will not contribute
to ${\cal O}(r^{-18})$.
Now, we can write out the $\Rcal^4$ term in the Lagrangian and find
\be
 c t_8 t_8 \Rcal^4 = c\cdot (2^2)^4(-\frac{1}{2})^4 \ t^{\mu_1 \cdots \mu_8}_8
   \  t^{\nu_1 \cdots \nu_8}_8 \ \pat_{\mu_1}\pat_{\nu_1}\Delta_{\mu_2\nu_2}
\cdots \pat_{\mu_7}\pat_{\nu_7}\Delta_{\mu_8\nu_8} + \cdots
\label{e17}
\ee
Then, substituting (\ref{e14}) and keeping terms with 2 factors 
of $\pat \pat h$ gives
\be
  c t_8t_8 \Rcal^4 = c\ \kappa^2 \cdot 6 \cdot  2^4
            \  t^{i-k-\mu_5\cdots \mu_8}_8 \ 
          t^{j-l-\nu_5\cdots \nu_8}_8\  \pat_i\pat_jh\ \pat_k\pat_lh
                           \ \pat_{\mu_5}\pat_{\nu_5}f_{\mu_6\nu_6}
                           \pat_{\mu_7}\pat_{\nu_7}f_{\mu_8\nu_8} + \cdots \ .
\label{e18}
\ee
Note that (\ref{e18}) shows explicitly that the effect we are calculating
is polarization dependent\footnote{See also the discussion in Appendix A.}. 
For the rest of the calculation we will
fix the polarization of the graviton to be in the 12-plane, as in (\ref{f12}).
Then, we find that
\be
   \sqrt{g} \frac{c}{\kappa^{2/3}} t_8t_8\Rcal^4 
      = c\ \kappa^{4/3} K^{\mu\nu\alpha\beta}
        \pat_{\mu}\pat_{\nu}f_{12} \pat_{\alpha}\pat_{\beta}f_{12}
\label{e19}
\ee
where we have used $\sqrt{g} = 1 + \cdots$ ,
since terms denoted by $+\cdots$ would only give terms that are higher
order in $h$ or $f_{12}$, and we have introduced the tensor 
notation $K^{\mu\nu\alpha\beta}$  for
\be
 K^{\mu\nu\alpha\beta} \equiv 6\cdot 2^5 \ \pat_i\pat_jh\ \pat_k\pat_lh
         \ \{t^{i-k-\mu 1\alpha 1}_8 \ t^{j-l-\nu 2 \beta 2}_8 
          +  t^{i-k-\mu 1\alpha 2}_8 \ t^{j-l-\nu 2 \beta 1}_8 \} \ .
\label{e20}
\ee   
Thus, the Lagrangian (\ref{e9}) becomes, to second order in $f_{12}$,
\be
 \Lcal = \frac{1}{4}\{ -2\pat_{\tau}f_{12}\pat_-f_{12} 
   - (\pat_{\perp}f_{12})^2 +h(\pat_{\tau}f_{12})^2 \}
   + c\ \kappa^{4/3}\  K^{\mu\nu\alpha\beta} \pat_{\mu}\pat_{\nu}f_{12}
          \pat_{\alpha}\pat_{\beta}f_{12} \ .
\label{e21}
\ee
In terms of the probe momenta, the field 
equation for $f_{12}$ is\footnote{Here we have 
used $\pat_{\mu}\pat_{\nu}K^{\mu\nu\alpha\beta} 
= K^{\mu\nu\alpha\beta}\pat_{\mu}\pat_{\nu} + \cdots $, where $+ \cdots $ again
denotes irrelevant terms.}
\be
  -2p_{\tau}p_- -p^2_{\perp} +hp^2_{\tau} 
    + 4c\ \kappa^{4/3}K^{\mu\nu\alpha\beta}p_{\mu}p_{\nu}p_{\alpha}p_{\beta}
    = 0 \ .
\label{e23}
\ee
Solving iteratively for $p_{\tau}$, we obtain
\be
 p_{\tau} = p^{(0)}_{\tau} + \frac{2c \ \kappa^{4/3}}{p_-} 
 \{ K^{\tau\tau\tau\tau} (p^{(0)}_{\tau})^4 
  + 4K^{\tau\tau\tau n} (p^{(0)})^3 p_n 
  + \cdots \}
\label{e24}
\ee
where $p^{(0)}_{\tau}$ is the solution (\ref{e7}) found previously, and
the index $n$ sums over $-$ and the transverse coordinates $i$.

The final task is to perform the Legendre transformation to find the Routhian,
and to restore the velocity $v_{\perp}$. Although all of the terms in the
curly brackets in (\ref{e24}) will produce $r^{-18}$ dependent terms,
only the first one contributes to a term of the form 
$v^8_{\perp} / r^{18}$, which has the velocity dependence we
choose to focus on.  In the Legendre transformation
it is sufficient to use the approximation
$p^{(0)}_{\tau} \approx -p^2_{\perp}/{2p_-} + {\cal O}(h)$
in the $(p^{(0)}_{\tau})^4$ term. After some algebra we
find the Routhian
\be
   L'
     =  \frac{p_-}{h} \{ 1- \sqrt{1-hv^2_{\perp}} \} 
    + \frac{c~\kappa^{4/3}}{8} K^{\tau\tau\tau\tau} p^3_- v^8_{\perp} 
     + \cdots  \ . 
\label{result}
\ee
The coefficient of the last term can be found explicitly.
Substituting (\ref{e10}), $p_- = N_p/R$, and from Appendix B,
\be
 K^{\tau\tau\tau\tau} = 6 \cdot 2^5 \cdot
\frac{7^2 4^2 15^2 N^2_s}{R^4M^{18}} \frac{1}{r^{18}}  
\ee
and
\be
  \kappa^{4/3} = \frac{4^{4/3}\pi^{10/3}}{M^6} \ ,
\ee
we finally find that the last term in $L'$ is
\be
        14700 \pi^5 \cdot ( 4\pi )^{1/3} 
     \cdot   \frac{N^2_sN^3_p}{M^{24}R^7}
         \frac{v^8_{\perp}}{r^{18}} \ .
\label{e25}
\ee
As discussed in the introduction, the contribution (\ref{e25}) has the
correct $v$ and $r$ dependence to match onto a two loop Matrix Theory term
in (\ref{lform}), but the $N_p$ dependence is of the wrong form.  

Although we have focussed on the $v^8/r^{18}$ contribution from 
the $\Rcal^4$ term,
there are other contributions as well whose forms we could easily derive.  
Clearly, by expanding $\sqrt{g}t_{8}t_{8}\Rcal^{4}$ in different ways one can
generate a wide variety of different contributions.

\section{Discussion}

It is useful to clarify, in the language of Feynman diagrams, which effects
our calculation is including and which  it is leaving out. In quantum 
mechanics language, we are doing a calculation of potential scattering,
in which one neglects the change in state of one of the scattered particles
(the source) and uses only the classical potential produced by that particle.
This is known as the effective field approximation; as discussed in 
\cite{Wein} it is valid when the source particle is heavy and non-relativistic.
In the case of gravitons in the light cone frame, this condition is 
$N_s \gg 1$ and $ p_{\perp} \ll p_{-}$.  

In our approach, we are writing the action as 
$$
  S(g^{\rm (cl)}_{\mu\nu}+f_{\mu\nu})
$$
and expanding to quadratic order in $f_{\mu\nu}$. 
Here $g^{\rm (cl)}_{\mu\nu}$ is a classical solution to the action 
\be
  S(g_{\mu\nu}) + \int d^{11}x \sqrt{g} g_{\mu\nu}T^{\mu\nu} 
\ee
where $T^{\mu\nu}$ is the energy-momentum tensor of the source graviton. We 
could evaluate the metric seen by the probe by calculating 
\be
 \langle g_{\mu\nu} \rangle = \int {\cal D}g \, g_{\mu\nu} 
 e^{iS(g_{\mu\nu})+i\int\! \sqrt{g}\,g_{\mu\nu}T^{\mu\nu}} 
\ee
at the location of the probe.
Keeping only the tree diagrams yields the classical metric,  
$\langle g_{\mu\nu} \rangle = g^{\rm (cl)}_{\mu\nu}$. 
Pictorially, we are keeping diagrams like Fig. (2a) but discarding loop
diagrams like Fig. (2b).

\def\fpsangle{0}
\begin{center}
\leavevmode
\fpsxsize 2.0in
\fpsysize 2.0in 
\fpsbox{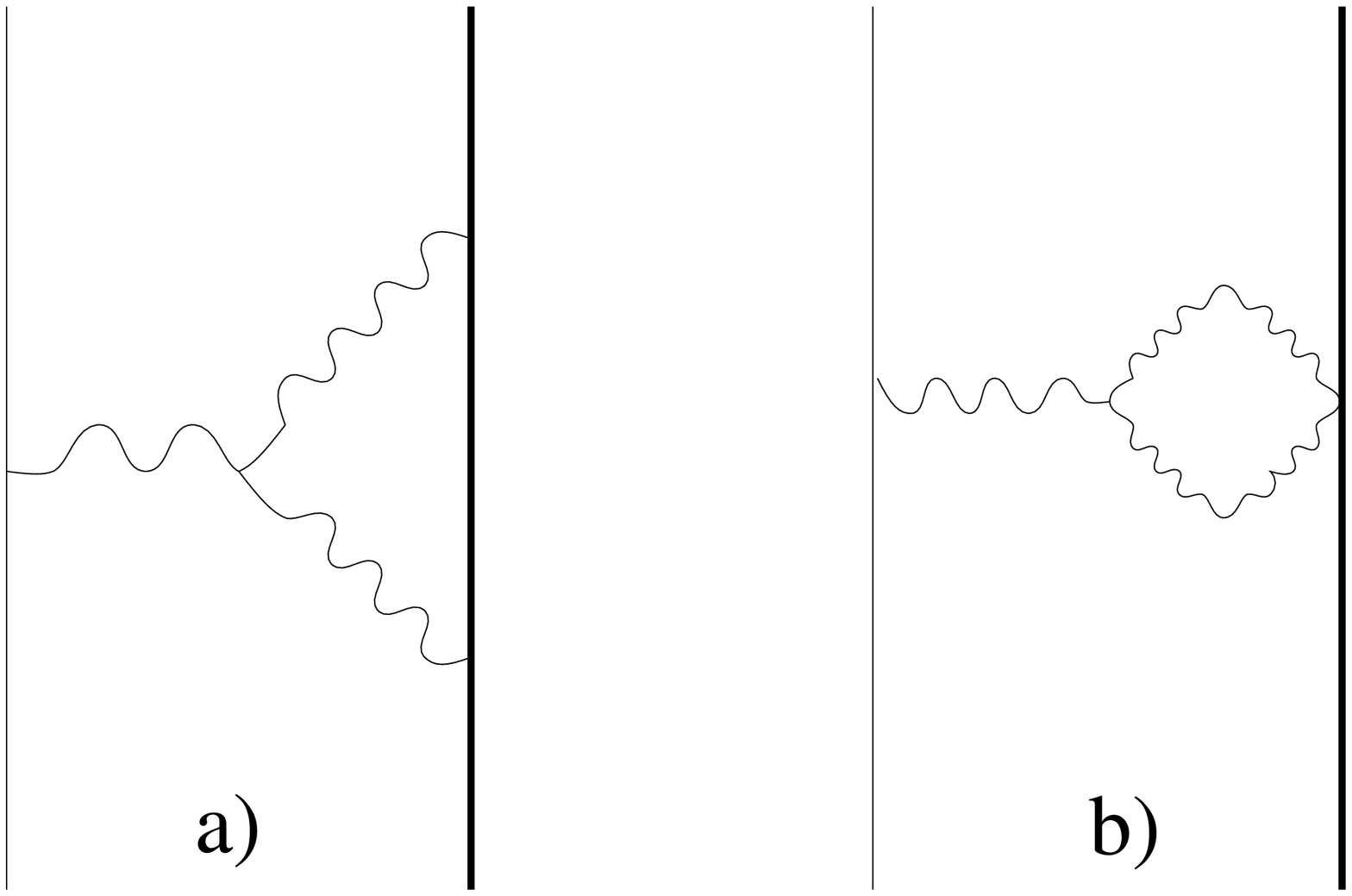}
\end{center}
\vskip 0.1 true cm
\begin{quotation}
\small
\noindent
{\bf Figure 2:}
a) contribution corresponding to the classical metric of the source; 
b) quantum correction to the metric.
\end{quotation}

In treating the source graviton as a fixed classical metric we are neglecting 
recoil effects. This is valid provided that the emission of quanta by the 
source does not change its state appreciably. In other words, we require that 
the momentum exchanged in the scattering process should be small in 
comparison to the momentum of the source graviton. 

In using only the 
classical value for the source metric $\langle g_{\mu\nu} \rangle$ we are 
neglecting certain correlation effects. That is, our calculation produces 
\be
 \langle g_{\mu\nu}(x)g_{\alpha\beta}(y) \rangle
 = \langle g_{\mu\nu} (x) \rangle \langle g_{\alpha\beta} (y) \rangle \ ,
\ee
and does not include corrections to this relation due to diagrams like
that in Fig. 3.

\begin{center}
\leavevmode
\fpsxsize 0.8in
\fpsysize 0.8in
\fpsbox{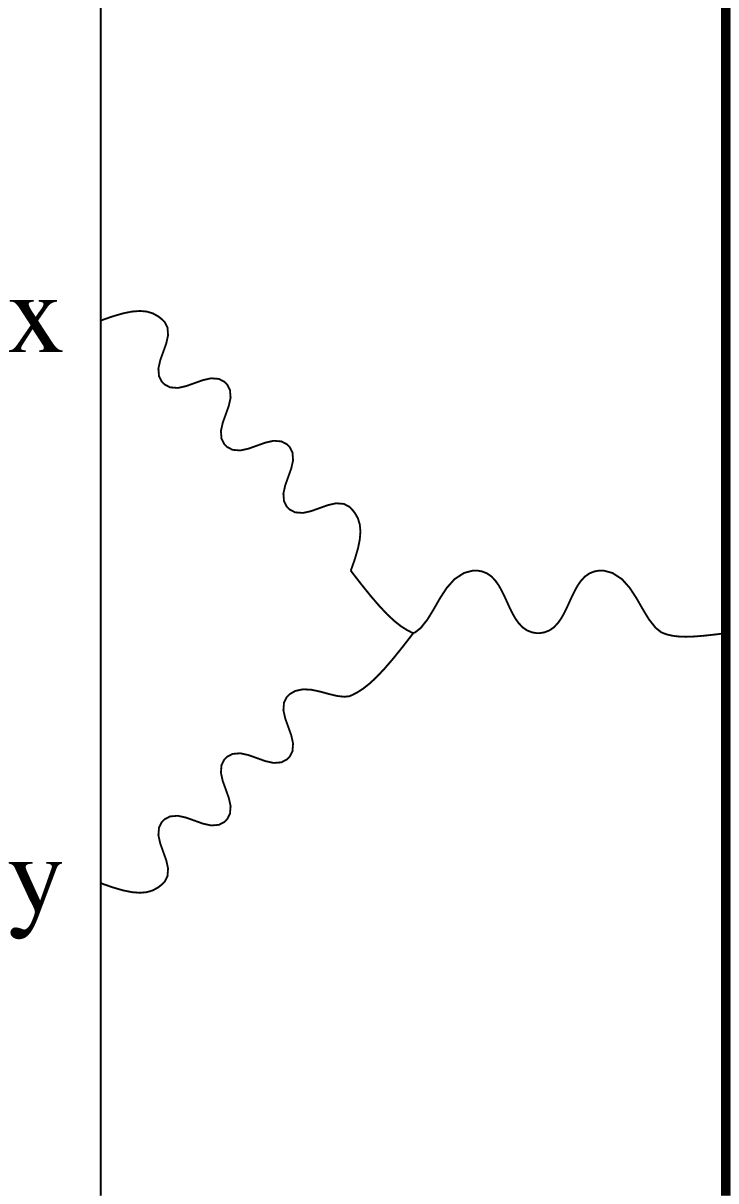}
\end{center}
\vskip -0.1 true cm
\begin{quotation}
\small
\nopagebreak
\noindent
{\bf Figure 3:}
Diagram showing correlation of source fields on the probe.
\end{quotation}
Such processes correspond to non-local terms in the effective action for the 
probe and are not captured by our analysis. 

Nowhere in our calculation did the anti-symmetric tensor or gravitino appear,
and so one might be led to consider the possibility that the undesirable
contribution we have found could be cancelled against diagrams involving
the exchange of these fields.  To see that this cannot happen, let us recall
that the $\Rcal^4$ arises from the divergent part of one loop diagrams,
including the ones where the anti-symmetric tensor and gravitino circulate 
in the loop (see Fig. 1).  Thus, at the level of two graviton exchange 
diagrams, 
we have already taken these fields into account insofar
as their divergent effects are concerned. The only remaining contributions
are the UV finite parts, but these have the property of being polynomial
in $\kappa$, since non-polynomial dependence can only arise through the
appearance of the UV cutoff.  When we note that our result (\ref{sugres}) 
contains a fractional power of $\kappa$, we conclude 
that it cannot be cancelled by the UV finite parts.  
Finally, in addition to $\Rcal^4$ there are $\Rcal^5$ and higher operators, 
but these 
can contribute only at the level of three or more graviton exchange diagrams,
and so cannot yield a $r^{-18}$ dependence. 

\section{Conclusion}

We have seen that the presence of the higher 
dimension $\Rcal^4$ operator in the
low energy supergravity action leads to terms in the graviton-graviton
scattering amplitude which have no simple analogues in Matrix Theory.
The most likely resolution of the discrepancy  is that 
we have been too naive in calculating
within the context of supergravity with lightlike compactification.  As
discussed recently in \cite{HP}, the presence of zero modes can 
complicate the dynamics of the lightlike compactified theory. In this case
case, our supergravity results would not be valid in the domain in which
they can be compared to finite $N$ Matrix Theory; instead, one should be
comparing with the $N\rightarrow \infty$ limit.  But in this limit one can
conceive of effects which would invalidate  the simple use of the effective
action in the way that is normally done.  It would then be important to
determine which processes in Matrix Theory are protected from receiving
$N\rightarrow \infty$ corrections so that they can be computed reliably
at finite $N$.

\bigskip

{\Large {\bf Acknowledgements}}

\bigskip

We would like to thank Petr Ho\v{r}ava and John Schwarz for useful discussions
and Senarath de Alwis for useful
comments on an earlier version of this paper.
The effect of the $\Rcal^4$ operator has also been studied by Katrin and
Melanie Becker \cite{BBnew}. 

\bigskip

\appendix{{\Large {\bf Appendix A}}}

\bigskip

Here make some comments on the polarization dependence of graviton scattering.
Initially, one might worry about terms in the equation of
motion for $f_{\mu\nu}$ which couple together different polarizations.
However, the polarization dependent effects  involve derivatives of $h$ and 
hence are subleading in $1/r$. This observation is seen from the 
following argument.
In flat space, the polarization does not affect the motion of the
graviton. In curved space, one can always transform to locally flat 
coordinates, where space is flat up to subleading corrections. 
The polarization dependence must then come from the subleading terms.
The transformation to locally flat coordinates, appropriate to
the metric (\ref{bg}), is simply 
$\tau \mapsto \tilde{\tau} = \tau + hx^-/2$, so the subleading corrections
and the polarization dependence arise from derivatives of $h$.

However, with the inclusion of the $\Rcal^4$ term the situation
is somewhat different. Here the point is precisely to study the subleading 
corrections depending
on derivatives of the harmonic function. Such corrections are expected to
be polarization dependent, which is indeed the case as one can 
see from (\ref{e18}).
In this paper we are only studying processes
where the polarization remains fixed, which are simpler to handle.

\bigskip
\appendix{{\Large {\bf Appendix B}}}
\bigskip
  
Here we provide some details of the calculation of  $K^{\tau\tau\tau\tau}$
which were skipped in section 4.
The first step is to find explicit formulas for the components of
$t_8$ which appear in the definition of $K^{\tau\tau\tau\tau}$. We proceeded
as follows. First, we completed the antisymmetrization
of $t_8$ as outlined in \cite{GSW}. This results in a long expression
with 60 terms, each involving a product of 
four metric factors $\eta^{\mu\nu}$.  
However, using 
$$
 \eta^{-\tau} \neq 0\ ; \ \eta^{ii} \neq 0 \ ; \ {\rm others} =0
$$
we get a much more compact
equation for the following components of $t_8$ which appear in the 
definition of $K^{\mu\nu\alpha\beta}$:
\begin{eqnarray}
 t^{i-k-\mu_5\cdots \mu_8}_8 &=& \eta^{ik }
                 \{ \eta^{-\mu_5}(\eta^{\mu_6\mu_8}\eta^{\mu_7-}
                          -\eta^{\mu_6\mu_7}\eta^{\mu_8-}) \nonumber \\
 & & \mbox{} \ \  -\eta^{-\mu_6}(\eta^{\mu_5\mu_8}\eta^{\mu_7-}
                         -\eta^{\mu_5\mu_7}\eta^{\mu_8-})\} \ \ .
\label{a1}
\end{eqnarray}
Now we can evaluate $K^{\tau\tau\tau\tau}$. Recall that we defined
$K^{\mu\nu\alpha\beta}$ by (\ref{e20}). Thus, the 
component $K^{\tau\tau\tau\tau}$ is given by
\be
 K^{\tau\tau\tau\tau} = 6\cdot 2^5 \ 
          \pat_i\pat_jh\ \pat_k\pat_lh \ 
  \{ t^{i-k-\tau 1\tau 1}_8 t^{j-l-\tau 2\tau 2}_8
   + t^{i-k-\tau 1\tau 2}_8 t^{j-l-\tau 2\tau 1}_8 \} \ .
\label{a2}
\ee
Using the result (\ref{a1}) above, we get
\begin{eqnarray}
   t^{i-k-\tau 1\tau 1}_8 &=& \delta^{ik}(\eta^{-\tau})^2
                                   \nonumber \\
   t^{j-l-\tau 2\tau 2}_8 &=& \delta^{jl}(\eta^{-\tau})^2 \ .
\end{eqnarray}
Similarly,
\be
  t^{i-k-\tau 1\tau 2}_8 = t^{j-l-\tau 2\tau 1}_8 = 0 \ .
\ee
Substituting to (\ref{a2}), we find 
\begin{eqnarray}
 K^{\tau\tau\tau\tau} 
        &=& 6\cdot 2^5 \cdot \pat_i\pat_jh \ \pat_i\pat_jh \nonumber \\
        &=& 6\cdot 2^5 \cdot 
         \frac{7^2 4^2 15^2 N^2_s}{R^4M^{18}}\frac{1}{r^{18}} \ .
\end{eqnarray}
In the last line, we substituted (\ref{e3}).

\end{document}